\documentclass[12pt]{article}
\usepackage[margin=2cm]{geometry}
\usepackage{comment}
\usepackage{amsmath,amssymb,extarrows,mathtools,graphicx,subfigure,setspace}
\usepackage{cite}
\usepackage{slashed}
\usepackage{color}
\makeatother

\def\BC{\mathbb C}
\def\_\BC{\mathbbi C}

\def\R{\mathbb{R}}
\def\N{\mathbb{N}}

\def\Z {\mathbb{Z}}

\def\bee{\begin{enumerate}}
\def\eee{\end{enumerate}}

\newcommand{\bei}{\begin{itemize}}
\newcommand{\eni}{\end{itemize}}

\newcommand\ben{\begin{enumerate}}
\newcommand\een{\end{enumerate}}

\def\R{{\rm I\hspace{-.15em}R}}
\numberwithin{equation}{section}
\begin{document}
\onehalfspacing
\vfill
\begin{titlepage}
\vspace{10mm}
\begin{flushright}
 IPM/P-2014/002 \\
\end{flushright}
\begin{center}
{\Large {\bf Particle Creation in Global de Sitter Space: Bulk Space Consideration}\\
}

\vspace*{15mm}
\vspace*{1mm}
{M. Reza Tanhayi}

 \vspace*{1cm}

{\it  Department of Physics, Islamic Azad University, Central Tehran Branch, Tehran,
 Iran\\School of Physics, Institute for Research in Fundamental Sciences (IPM)\\
P.O. Box 19395-5531, Tehran, Iran\\
}
 \vspace*{0.5cm}
{E-mail: {\tt mtanhayi@ipm.ir}}%

\vspace*{1cm}
\end{center}

\begin{abstract}

Recently in \cite{Anderson:2013zia, Anderson:2013ila}, it was shown that global de Sitter space is unstable even to the massive particle creation with no self-interactions. In this paper we study the instability by making use of the coordinate-independent plane wave in de Sitter space, within this formalism, we show that the previous results of instability of de Sitter space due to the particle creation can be generalized to higher-spin fields in a straightforward way. The so-called plane wave are defined globally in de Sitter space and de Sitter invariance is manifest since such modes are deduced from the group theoretical point of view by the means of the Casimir operators. In fact, we employ the underling symmetry of embedding space namely the $4+1$ dimensional flat space to write the field equations and the solutions can be obtained in terms of the plane wave in embedding space. 
\end{abstract}
\end{titlepage}

\section{Introduction}

The study of particle creation in expanding universe has a rich history dated back to the early work of Schr\"odinger in 1939 and followed by Parker in the late 1960s \cite{parker}. Basically, if one could define positive and negative frequency solutions at the future and past, separately, the particle creation can be understood by the off-diagonal elements of the matrix which relates these two sets of solution via the linear transformation. The de Sitter space-time is of a particular importance which has been considered in the literature. Cosmological observations \cite{Riess} strongly suggest that the evaluation of our universe is dominated by a positive and non-vanishing cosmological constant, it may be viewed as a vacuum energy and thus the cosmological constant can be responsible of the measured acceleration of the universe \cite{mukh}. This means that  de Sitter space can well approximate the universe compared to the other known models. The de Sitter space also plays an important role in explanation of the inflationary model at early universe. The cosmological constant is related to the curvature of space-time through the constant Ricci scalar as $R=4\Lambda$ and the corresponding
fundamental length is given by $H^{-1}=
\sqrt{3/\Lambda}$, where $H$ is the Hubble constant.\footnote{Note that $\hslash=1=c$ are used.}  The de Sitter space together with anti-de Sitter space are the maximally symmetric solution to Einstein's equations and hence, the underling symmetries can be used to construct the field theory on this background, studied in several papers \cite{dirac, Allen, Allen2, borner, bros, bros1, tagi, Bousso:2001mw, takook1}. The symmetries can be encoded according to de Sitter group namely the ten-parameter $SO_0(1,4)$ which is a deformation of the kinematical group of Minkowski space-time or the proper orthochronous Poincar\'e group.  Here, we follow this approach to obtain the field equations for arbitrary spin in de Sitter background.  

Actually, in de Sitter space one can not introduced a well-defined Hamiltonian and hence its ground state as a vacuum state is not clear \cite{brill}. Basically, in Minkowski space-time thanks to the Poincar\'e group, the Hamiltonian with the time translational symmetry can be defined and therefore, two different types of positive and negative spectrum exist which are not allowed to mix. Thus Hamiltonian's ground state can be regarded as a true symmetric vacuum state for all inertial Poincar\'e invariant observers. Generally, in time dependent curved spacetimes including de Sitter space, there is no time-like Killing vector and consequently, the definition of the vacuum and particles are problematic. In other words, the action of the de Sitter group generators allows the transformation positive and negative frequency solutions and therefore, the stability of de Sitter states under the particle creation is under debate \cite{poly, Anderson:2013ila}.  Nevertheless, in the de Sitter space instead of defining Hamiltonian, the maximally symmetric vacuum state or the Bunch-Davies (Euclidean) vacuum  is usually used which first introduced in \cite{tagi}. The particle creation is understood by the means of the Bogoliubov coefficients between the $in$ and $out$ vacua, where the $in$ (or similarly $out$) vacuum state is the vacuum for the fields coming from the past (future) infinity. It is also worth mentioning that in a time varying background such as de Sitter space the definition of a particle is intrinsically non-unique. However, in Ref. \cite{parker1} it was argued that the identification of physical particles becomes exact only in the limit of adiabatic (slow) expansion which this is achieved by setting an upper limit on the creation rate in the present expanding universe. To carry out, a higher-order WKB approximation to the time dependence of the scalar was developed through an iterative procedure called successive adiabatic approximation. Also in Ref. \cite{habib} the adiabatic particle concept which based on adiabatic basis was used to address non-uniqueness considerably. In this way it has been argued that one must use fourth order adiabatic states and these physical states are UV allowed Robertson-Walker state with short distance behavior consistent with general covariance of the low energy effective field theory and the Equivalence Principle.\footnote{Note that in the Robertson-Walker space-time, the energy-momentum tensor is conserved mode by mode for any state the field is in, not just fourth order adiabatic states so long as adiabatic regularization is used. However, for an arbitrarily coupled scalar field in a general Robertson-Walker space-time, lower order adiabatic states does not work. Actually the main problem with lower order adiabatic states is that in general the energy-momentum tensor diverges.  This is almost certainly related to the violation of the equivalence principle.  In fact the divergences in the energy-momentum tensor occur when the space-time points come together.} In the case of de Sitter space, the O(1,4)-invariant Bunch-Davies state is a UV allowed fourth order adiabatic state (for more details see \cite{Anderson:2005hi} and references therein).
 
Having noticed that in de Sitter space the concept of particle and positive energy solution are not well-defined invariant parameter for a set of observers transforming under the de Sitter group $SO(1,4)$, however in this paper, from the group theoretical point of view, the massive (massless) particle refer to those unitary irreducible representation of de Sitter group which coincide to the massive (massless) representation of the Poincar\'e group at the zero curvature limit (Minkowskian limit) \cite{Gazeau1}. 

There are a lot of papers considering the various features of particle production in de Sitter space, (see for example \cite{mottela} and references therein). Recently in Ref.s \cite{Anderson:2013zia, Anderson:2013ila}, the instability of de Sitter space under the particle creation has been discussed, here we use the group theoretical approach which is well established in the plane wave formalism to analyse the question of instability of de Sitter under particle production and this approach could potentially generalize to higher-spin fields, which are discussed. In order to define the fields in de Sitter space, we use the analytic continuation of the plane wave introduced in \cite{bros, bros1}. From the analyticity in the complexified Riemannian manifold and by the extension of the Fourier-Helgason transformation in de Sitter space, the coordinate-independent plane wave in de Sitter space was introduced. The so-called plane wave gives rise to the thermal interpretation for the Bunch-Davies vacuum in de Sitter space and as it is shown at the zero-curvature limit, this wave coincides to the usual plane wave in Minkowski space. The scalar plane waves is given by
\begin{equation}\label{plane}
\phi(x,\xi)={\cal N}(x\cdot\xi)^\sigma,
\end{equation} 
 where $x$ is the embedding coordinate, ${\cal N}$ and $\sigma$ are the normalization factor and homogeneity degree respectively. The null five-vector, $\xi$, is introduced which plays the role of the $4$-momentum in flat space in order to label the plane waves in de Sitter space, we will get back to this latter.  In this formalism, the correlation functions are
boundary values of analytical functions and the analyticity condition is only preserved in the case
of Euclidean vacuum. Making use of ambient de Sitter space and plane wave method has some other advantages, for example, in \cite{QNM} it is shown that one can reformulate the quasi-normal modes in de Sitter space in an easier way. 

The layout of the paper is as follows: In section 2, the field equations for various 'spins' are studied in the context of the plane wave in the ambient space. The solutions are found in terms of the de Sitter plane wave in section 3, and it is shown that other higher-spin fields can be written in terms of the scalar plane wave in a straightforward way. In section 4, we first recall that the plane wave could be defined in whole de Sitter space and then we study the possible mixture of the positive and negative frequency solutions due to the such modes at the early time. Finally we have enclosed paper with a brief conclusion. Some mathematical relations are given in appendix.

\section{de Sitter Modes: Basic Set-up}
                
The de Sitter space is a maximally symmetric solution of the cosmological vacuum Einstein's equations with positive $\Lambda$. The de Sitter geometry can be seen as a $4$-dimensional hyperboloid embedded in $4+1$-dimensional (bulk) Minkowski space-time

\begin{equation}
X_H=\{x \in \R ^5;\,\,x^2=\eta_{\alpha\beta} x^\alpha x^\beta
=-H^{-2}=-\frac{3}{\Lambda}\},\;\;\alpha,\beta=0,1,2,3,4,
\end{equation}
where $\eta_{\alpha\beta}= diag(1,-1,-1,-1,-1)$. 
The $X^\mu$'s are the intrinsic coordinates. Any geometrical object in this space can be
written in terms of $X^\mu$ or in terms of the five global bulk coordinates $x^\alpha$ (ambient space). 

\subsection{Field Equations: Group Theoretical Approach} 

The isometry group of de Sitter is the ten-parameter homogeneous Lorentz group $SO_0(1,4)$ with two coordinate invariant Casimir operators. The fields can be classified according to the irreducible representations of the de Sitter group which are labelled by the eigenvalues
associated with the Casimir operators. In fact, only one of the Casimir operators is relevant in characterizing fields since it leads to the quadratic field equations which it is given by:
\begin{equation}
Q=-\frac{1}{2}L_{\alpha\beta}L^{\alpha\beta},
 \end{equation}
where $L_{\alpha\beta}=M_{\alpha\beta}+S_{\alpha\beta}$, the orbital part is defined by $M_{\alpha\beta}=-i(x_{\alpha}\partial_{\beta}-x_{\beta}\partial_{\alpha})$ while the spinorial part $S_{\alpha\beta}$ acts
on the indices of a given function in a certain way, for example in the case of rank 2 tensor field one has:
$$ S_{\alpha\beta}{\cal K}_{\gamma\delta}=-i(\eta_{\alpha\gamma}{\cal K}_{\beta\delta}-\eta_{\beta\gamma} {\cal K}_{\alpha\delta} +
 \eta_{\alpha\delta}{\cal K}_{\beta\gamma}-\eta_{\beta\delta}
 {\cal K}_{\alpha\gamma}),$$
and in the spinoral case, we have \cite{clas}: 
\begin{equation}
S^{(1/2)}_{\alpha\beta}=\frac{-i}{4}[\gamma_\alpha,\gamma_\beta],
\end{equation}
note that here, one needs five $4\times4$ $\gamma_\alpha$  matrices which are the generators of the Clifford algebra based on the metric $\eta_{\alpha\beta}$, namely: 

\begin{equation}
\gamma^\alpha\gamma^\beta-\gamma^\beta\gamma^\alpha=2\eta^{\alpha\beta}\mathbb{I},\,\,\,\,\gamma^{\alpha\dagger}=\gamma^0\gamma^\alpha\gamma^0,
\end{equation}
where $\mathbb{I}$ is the identity operator. In the case of spinor fields, $s=l+\frac{1}{2}$, one can write: $S^{(s)}_{\alpha\beta}=S_{\alpha\beta}+S^{(1/2)}_{\alpha\beta}$.
  
Here, we are only interested in representations of de Sitter group that reduce to the massive representations of the Poincar\'e group in the flat limit. They are named as the Principal series, $U_{\nu,s}$ where $s$ stands for the spin \cite{Dix, gra} and the proper eigenvalues are given by
\begin{equation}
\label{q1} 
<Q> = \big(\tfrac{9}{4} + \nu^2 -s(s+1) \big)\mathbb{I} ,
\end{equation}
noting that  $\nu \in \R$, if $s=1,2, \dots$, and  $\nu \neq 0$, for $s= \frac{1}{2}, \frac{3}{2}, \frac{5}{2}, \dots$. It is proved that in taking flat limit these representations reduce to the Poincar\'e  representations as illustrated in the following diagram:
\begin{gather*}
U_{\nu,s} \underset{H \to 0}{\longrightarrow} {\cal P}^{>}(m,s)
\oplus {\cal P}^{<}(m,s),
\end{gather*}
where ${\cal P}^{\stackrel{>}{<}}(m,s)$ denotes
the positive (resp.\ negative) energy UIRs of the Poincar\'e  group with mass $m$ and spin $s$. Therefore, the second order wave equation can be written as:
\begin{equation}
\label{q}
(Q^{(s)}-<Q^{(s)}>)\psi=0.
\end{equation}
This means that one may consider the eigenvector equations of the second-order Casimir operator which for any eigenvalue, they give a Klein-Gordon-like or Dirac-like equation. From (\ref{q1}) and (\ref{q}), the field equations for a field of spin $s$ in de Sitter space may be written as: 
\begin{equation}
\label{scalar}
\Big(Q^{(s)}-(\nu^2+\frac{9}{4}-s(s+1))\Big)\psi(x)=0,
\end{equation}
where $\nu>0$ for the spinoral case and $\nu\geq 0$ for $s=0,1,2,\cdots$. 
For example in case of a scalar field the Eq. (\ref{scalar}) turns to: 
\begin{equation}
\label{scalarcas}
\Big(Q^{(0)}-(\nu^2+\frac{9}{4})\Big)\phi(x)=0, 
\end{equation} 
this equation has been written in terms of the Casimir operator, in this notation, the relationship with unitary irreducible representation of the de Sitter group becomes straightforward because the Casimir operators are easily identified with the field equation. The obtained equations are written in $4+1$ ambient de Sitter space, to obtain their corresponding $3+1$ de Sitter hyperboloid counterparts, the projection is needed. 

The transverse tensor field
$\phi_{\alpha\beta\cdots}(x)$ is locally determined by the intrinsic
field $\Phi_{\mu\nu\cdots}(X)$ through 
\begin{equation}\label{passage}
\Phi_{\mu\nu\cdots}(X)=\Big(\frac{\partial x^{\alpha}}{\partial
X^{\mu}}\frac{\partial x^{\beta}}{\partial
X^{\nu}}\cdots\Big)\phi_{\alpha\beta\cdots}(x(X)),
\end{equation}
where the $\phi_{\alpha\beta\cdots}(x)$ is supposed to be a transverse homogeneous function. Any tensor field can be translated into the ambient space notations via the transverse projection
\begin{equation}
(Trpr \phi)_{\alpha_1 \cdots \alpha_l}\equiv
\theta_{\alpha_1}^{\beta_1}
\cdots\theta_{\alpha_l}^{\beta_l}\phi_{\beta_1 \cdots \beta_l}\;\end{equation}
where the projection operator is given by
\begin{equation}
\theta_{\alpha\beta}=\eta_{\alpha\beta}+H^2x_\alpha x_\beta,
\end{equation}
As a matter of fact the transverse projection guarantees the transversality in each
index. Thus the covariant derivative of a tensor field, in the ambient space notation becomes
\begin{equation}
 Trpr \partial^T_\beta \phi_{\alpha_1 ..... \alpha_n}\equiv
\nabla_\beta \phi_{\alpha_1....\alpha_n}\equiv 
\partial^T_\beta
\phi_{\alpha_1....\alpha_n}-H^2\sum_{i=1}^{n}x_{\alpha_i}\phi_{\alpha_1..\alpha_{i-1}\beta\alpha_{i+1}..\alpha_n}.\end{equation}
In the above relations $\partial^T_{\alpha}$ is the tangential
(or transverse) derivative on de Sitter space,$$
\partial^T_{\alpha}=\theta_{\alpha\beta}\partial^{\beta}=\partial_{\alpha}+H^2x_{\alpha}x\cdot\partial,\,\,\,\,\,x\cdot\partial^T=0.$$
And also one can write: 
$$g^{dS}_{\mu\nu}=\frac{\partial x^\alpha}{\partial
X^\mu}\frac{\partial x^\beta}{\partial X^\nu}
\theta_{\alpha\beta},$$
$$ \nabla_\mu  \cdots \nabla_\rho \Phi_{\lambda_{1} \cdots
\lambda_{l}}=\frac{\partial x^\alpha}{\partial X^\mu} \cdots
\frac{\partial x^\gamma}{\partial X^\rho}\frac{\partial
x^{\eta_1}}{\partial X^{\lambda_1}}\cdots \frac{\partial
x^{\eta_l}}{\partial X^{\lambda_l}} Trpr \partial^T_\alpha
\cdots Trpr \partial^T_\gamma \phi_{\eta_1 \cdots \eta_l}.$$ 
Therefore it is easy to show that the Laplace-Beltrami operator $\Box_H$, on de Sitter manifold turns to 
\begin{equation}
\Box_H\Phi(X)=g^{dS}_{\mu\nu}\nabla^\mu\nabla^\nu\Phi(X)
\rightarrow\eta^{\alpha\beta}\partial_\alpha^T\partial_\beta^T\phi(x)
\end{equation}
 $g^{dS}_{\mu\nu}$ is the de Sitter metric and $\nabla_\mu$ is the covariant derivative in $X$ coordinates. On the other hand one can show that:
 \begin{equation}
 \begin{array}{l}
 Q^{(0)}\equiv-\frac{1}{2}M_{\alpha\beta}M^{\alpha\beta}\\
 =-\frac{1}{2}\Big(x_\alpha\partial^T_\beta-x_\beta\partial^T_\alpha\Big)\Big(x^\alpha\partial^{T\beta}-x^\beta\partial^{T\alpha}\Big)=-H^{-2}\Box_H
\end{array}  
  \end{equation}
As one knows the Klein-Gordon equation on the hyperboloid reads as\footnote{The Klein-Gordon equation comes from the variation principle from the following action:
$${\cal S}=\frac{1}{2}\int_{X_H}d^4X\sqrt{-g}\Big(g^{\mu\nu}\partial_\mu\phi\partial_\nu\phi-(M_H^2+\zeta R)\phi^2\Big)$$ where $\zeta$ is a positive
gravitational coupling with the de Sitter background, in this paper we consider the minimal coupling case. We review the asymptotic behaviors of the solutions at the appendix.} 
\begin{equation}
\label{scalarfi}
(\Box_H+M_H^2)\Phi(X)=0,
\end{equation}
 comparing with (\ref{scalarcas}), leads one to deduce that the mass parameter is 
 \begin{equation}
 M^2_H\equiv H^2(\nu^2+\frac{9}{4}).
 \end{equation}
This field equation is usually considered in the literature and hence the solutions are found in terms of the intrinsic variables in different coordinates.

\section{de Sitter Plane Wave: the Solutions}

For a scalar field satisfying the Klein-Gordon like wave equation in de Sitter background with a coupling constant, the solutions are obtained in different coordinates by separating the variables.  However, in this paper we follow the plane wave method, the so-called plane wave is the solution of (\ref{scalarcas}) in embedding de Sitter space. It is shown that in the zero curvature limit, it covers the usual plane wave in flat space and more importantly, it is a stepping stone for higher spin field solutions. As mentioned, there is no well-defined momentum in de Sitter space, it is actually needed to write a plane wave. In embedding de Sitter space, instead of four-momentum, a five vector $\xi^\alpha=(\xi^0,\overrightarrow{\xi},\xi^4)$ with $\xi^\alpha\xi_\alpha=\xi^2=0$ maybe defined which it plays the role of four momentum in the null curvature. Here, we review the plane wave shortly. On $X_H$ let us define the following set of points as
\begin{equation}\label{v}
V^+=\{ x\in \R^5, \eta_{\alpha\beta}x^\alpha x^\beta>0,\,\,\,x^0>0\}, 
\end{equation}
for any given two events $x$ and $x'$, we say that $x>x'$ or $x$ is in the future of $x'$, if $x-x'$ be a vector in $V^+$, similarly, one can define $V^-$ and consequently the past causal of $x'$ is defined events say as $x$ in which $x-x'\in V^-$. In other words, for a given $x\in X_H$, a causal future is defined by $\{x'\in X_H, x'\geq x\}$, and $V^\pm$ is called the future or forward cone \cite{bros}. Two points are in causal relation if they belong to the intersection of the complements of such sets  \cite{bros, Gazeau1}, and the causality may be understood by defining the future (or past) light cone $C^+$ which is the boundary of the forward (or backward) cone $V^+$ (or $V^-$):
\begin{equation}
C^+=\{x\in \R^5, \eta_{\alpha\beta}x^\alpha x^\beta=0, x^0\geq 0\}.\nonumber
\end{equation}
For $\xi\in C^+$, it is proved that there exists a continuous family of simple solutions to (\ref{scalarcas}) which is named as a plane wave solution, it reads as
\begin{equation}
\label{plane}
\phi(x)=\big(H x(X)\cdot \xi\big)^\sigma,
\end{equation} 
where $\sigma$ is a complex number, and hence, $C^+$ may be also interpreted as the space of (asymptotic) momenta directions. One can check that
\begin{equation}
\Big(\Box_H- H^2\sigma(\sigma+3)\Big)(H x\cdot \xi)^\sigma=0,
\end{equation} 
with $\nu^2+\frac{9}{4}=\sigma(\sigma+3)=\frac{M^2}{H^2}$, that means 
\begin{equation}
\sigma_\pm=-\frac{3}{2}\pm i\nu.
\end{equation}  Notably, with changing $\sigma\rightarrow -\sigma-3$ the mass term does not change and in what follows we take $\sigma=-\frac{3}{2}-i\nu$. 

We would like to mention that, by specific parametrizing of $\xi$ and $m$, the introduced plane wave (\ref{plane}) at the flat limit when $H\rightarrow 0$, reduces to the Minkowskian plane wave (Appendix)
\begin{equation}
\lim_{H\rightarrow 0} (H x\cdot \xi)^\sigma \longrightarrow e^{\pm i\textbf{k}\cdot X}.
\end{equation}
By making use of the plane wave, one may define the Fourier transformation in de Sitter space:
\begin{equation}
\tilde{f}(\xi,\sigma)=\int_{X_H}(Hx\cdot\xi) ^\sigma f(x)\,\,dx.
\end{equation} 
The other fields can be obtained in terms of this plane wave in which the de Sitter plane wave formalism offers a solution as:
\begin{equation}
\psi(x)={\cal D}(\xi,x)(Hx\cdot\xi)^\sigma,
\end{equation}
${\cal D}(\xi,x)$ is a vector-valued differential operator, accordingly, the quantum theory of fields which are based on, can be constructed in the de Sitter space. Here, we recall massive vector and tensor fields in de Sitter space. 

\subsection{Massive vector field} 

Form equation (\ref{scalar}) it is understood that the field equations become
\begin{equation}
\big(Q^{(1)}-(\nu^2+\frac{1}{4})\big){\cal A}_\alpha(x)=0.
\end{equation}
After doing some algebra one finds
\begin{equation}
Q^{(1)}
{\cal A}_{\alpha}=(Q-2){\cal A}_{\alpha}+2x_{\alpha}\partial^T \cdot {\cal A}
-2\partial^T_{\alpha} x\cdot {\cal A}, 
\end{equation}
and the solutions are found as \cite{Gazeau:1999xn} 
 \begin{eqnarray}
{\cal A}_\alpha(x,\xi)={\cal D}_\alpha^{\lambda}(x,\xi)(Hx\cdot\xi)^{-\frac{3}{2}-i\nu},
\end{eqnarray}
 together with its complex conjugate form two independent solutions, noting that in this case ${\cal D}_\alpha^{\lambda}(x,\xi)$ are the generalized polarization vector
 \begin{equation}
 {\cal D}_\alpha^{(\lambda)}(x,\xi)=(\frac{3}{2}+i\nu)\Big(
 Z^{T(\lambda)}_\alpha-(i\nu+\frac{1}{2})\frac{Z^{(\lambda)}\cdot x}{x\cdot \xi}\xi^T_\alpha\Big),
 \end{equation}
where $Z_\alpha$ is a constant five vector in embedding space. Noting that at the zero curvature limit the solution tends to 
\begin{equation}
\lim_{H\rightarrow0}{\cal A}_\alpha(x)=\epsilon_\mu^{(\lambda)}(k)e^{ik\cdot X},\hspace{3mm}\lambda=1,2,3
\end{equation}
where $\epsilon_\mu^{(\lambda)}(k)$ are the three polarization vectors in Minkowski space.

\subsection{Massive tensor field}

It is proved that by making use the plane wave method, the de Sitter tensor modes can be written straightforwardly in terms of plane wave and a generalized polarization tensor. The proper field equation is obtained from (\ref{scalar}) as
\begin{equation}
\big(Q^{(2)}-(\nu^2-\frac{15}{4})\big){\cal H}_{\alpha\beta}=0.
\end{equation} 
One can write this field equation in terms of scalar Casimir operator after making use of the following relation
\begin{equation} Q^{(2)} {\cal H}_{\alpha
\beta}=(Q-6){\cal H}_{\alpha \beta}+2{\cal
S}x_{\alpha}\partial^T\cdot {\cal H}_{\beta}-2{\cal
S}\partial^T_{\alpha}x\cdot {\cal
H}_{\beta}+2\eta_{\alpha\beta}{\cal H}',
\end{equation}
 where ${\cal S}$ is the symmetrizer operator acting on the indices and ${\cal H}'$ is the trace of tensor field. Imposing certain conditions on tensor field, e.g.,  ${\cal H}_{\alpha\beta}={\cal H}_{\beta\alpha}$, transversality $x\cdot{\cal H}_\beta=0$, which together with $\partial \cdot {\cal H}={\cal H}'=0$, reduces 25 components of ${\cal H}_{\alpha\beta}$ to 5 independent components which correspond to what expected for spin-2 field degree's of freedom. There are two independent solutions as 

 \begin{equation}
{\cal H}_{\alpha\beta}={\cal D}_{\alpha\beta}(x,\xi)(Hx\cdot\xi)^{-\frac{3}{2}\mp i\nu},
\end{equation} 
 where as for vector field, ${\cal D}_{\alpha\beta}(x,\xi)$ is a generalized polarization tensor which is a space-time function. It is given by
 \begin{equation}
{\cal D}_{\alpha\beta}=c_\nu(\frac{\sigma-1}{\sigma+1})[{\cal S}\epsilon_\alpha^\lambda\epsilon_\beta^{\lambda'}-\frac{2}{3}(\theta_{\alpha\beta}-\frac{\xi_\alpha^T\xi^T_\beta}{(Hx\cdot \xi)^2}\epsilon^\lambda\cdot\epsilon^{\lambda'})],
 \end{equation}
where $c_\nu$ is the normalization constant, $\epsilon^\lambda_\alpha(x,\xi)=(Z_\alpha^\lambda-\frac{Z^\lambda\cdot x}{x\cdot \xi}\xi_\alpha)$, in which $Z$ is an arbitrary five-vector. Note that the arbitrariness in choosing of this tensor can be indeed fixed in such a way that, in the flat limit, the polarization tensor in Minkowski space-time should be recovered (for more mathematical  details see \cite{Garidi:2003bg})
\begin{equation}
\lim_{H\rightarrow 0}\epsilon^\lambda_\alpha(x,\xi)=\epsilon^\lambda_\mu(k).
\end{equation}
 As it is seen the scalar plane wave plays the crucial role in writing other modes in de Sitter ambient space. 
  
\section{Stability and Particle Creation } 

The introduced plane waves for the negative values of $\sigma$ when $x\cdot\xi=0$ become singular. To have globally defined modes in de Sitter space, a prescription was proposed in \cite{bros1} where based on the holomorphy properties and the complexified de Sitter space-time. In the complex de Sitter space-time 
 \begin{equation}
X_H^{(c)}=\{z=x+iy;\,\,\eta_{\alpha\beta}z^\alpha z^\beta=-H^{-2}\}=\{ (x,y)\in
\R^5\times  \R^5;\;\; x^2-y^2=-H^{-2},\; x.y=0\},
\end{equation}
let introduce the forward and backward tubes say as $T^\pm=
\R^5+iV^\pm$, noting that the
domain $V^\pm$ which has been defined in (\ref{v}) stems from the causal structure on $X_H$.  
Respective intersections with $X_H^{(c)}$ are 
 \begin{equation} {\cal T}^\pm=T^\pm\cap X_H^{(c)}=\{x+iy\in X_H^{(c)},\,\,\,y\in \pm V_+\},\end{equation}
which is called future and past tuboids. These are actually the analyticity domains of quantum fields in de Sitter space satisfying the positivity of spectrum of energy operators. Now one can say that when $z$ varies in ${\cal T}^+$ (or ${\cal T}^-$) and $\xi$ lies on the positive cone $ C^+$
$$\xi \in  C^+=\{ \xi \in  C; \; \xi^0>0 \},$$
the plane wave solutions are globally defined, since the imaginary
part of $(z.\xi)$ has  a fixed sign. The phase is chosen such that
 \begin{equation} \mbox{boundary value of} \; (z.\xi)^\sigma \mid_{x.\xi>0}>0.\end{equation}
In other words $(x\cdot\xi)^\sigma_\pm=\lim_{z\rightarrow x}(z\cdot\xi)^\sigma$ when ${z\in{\cal T}_\pm}$ are homogeneous distributions of degree $\sigma$ on $5$ dimensional flat space which should be restricted to the $4$ dimensional de Sitter space and the fields are holomorphic functions on ${\cal T}_+\cup{\cal T}_-$. The physical meaning of the holomorphy properties in axiomatic field theory in flat space is actually to provide the positivity of energy-momentum spectrum \cite{streater}, interestingly, in de Sitter space they lead to the thermal properties of the Bunch-Davies vacuum \cite{bros1}. Now we have a solution which has been written in embedding space with de Sitter constraint, in order to establish a manifest relation between this solution and ordinary solution of a scalar field in de Sitter space let us write the plane wave in global de Sitter coordinates. At the appendix it is shown that the plane wave in the global coordinate $(t, X):=(\sinh \tau,\,\,\cosh \tau \,\textbf{u})$, can be written as 
\begin{eqnarray}
\label{newmode}
(X\cdot \xi )^{\sigma} =  4\pi^2 \sum_{Llm}\frac{e^{ i\frac{\pi}{2}(\sigma-L)}}{\cosh \tau}
\frac{\Gamma(L-1-\sigma)}{\Gamma(-\sigma)}P_{-\frac{1}{2}-i\nu}^{-L}(- i\sinh\tau)Y_{Llm}(\textbf{u}_\xi)Y_{Llm}(\textbf{u}_X),\nonumber\\
\mbox{for}\,\,\,0< Im \,\,\tau<\pi.
\end{eqnarray}
where $L=1,2,\cdots$ and for $0<Im\,\,\tau<\pi$, the events belong to the forward tube ${\cal T}_+$ and by the means of analytic continuation, one could extend to $-\pi<Im\,\,\tau<0$ where the events take place in ${\cal T}_-$, noting that $\textbf{u}$'s are two unit vectors in $S^3$. The modes that we are interested in can be obtained by integrating out the $\xi$ dependency which are 
\begin{eqnarray}
\varphi_\sigma=\frac{e^{-i\frac{\pi}{2}\sigma}\Gamma(\frac{3}{2}+i\nu)}{4\pi^2}\int (\xi\cdot x)^\sigma Y_{LlM}(\textbf{u}_\xi)\,d\xi \equiv \chi_\sigma(\tau)Y_{Llm}(\textbf{u}_X),
\end{eqnarray}
where 
\begin{equation}\label{ar}
\chi_\sigma(\tau)= \frac{e^{-\frac{ i\pi}{2}L}\Gamma(L+\frac{1}{2}+i\nu)}{\cosh \tau}P_{-\frac{1}{2}-i\nu}^{-L}(-i\sinh \tau).
\end{equation}
The obtained modes together with their complex conjugate in each domain form a complete set and accordingly the vacuum state is defined by $a(\xi)|vac>=0$. 

On the other hand, if one started from the usual massive scalar field equation in de Sitter background, the solution was found to be 
\begin{equation}
\phi(\tau)={\cal N} (\cosh\tau)^{-1}\Big(P_{-\frac{1}{2}-i\nu}^{-L}(i\sinh \tau)+Q^{-L}_{-\frac{1}{2}-i\nu}(i\sinh\tau)\Big),  
\end{equation}
where $P$ and $Q$ are the first and second kind of Legendre functions and ${\cal N}=\frac{1}{\Gamma(-L+\frac{1}{2}+i\nu)}\sqrt{\frac{e^{-\pi\nu}}{\sinh\pi\nu}}$ is the normalization constant that can be deduced from the Wronskian condition or the Klein-Gordon inner product{\footnote{The Klein-Gordon inner product is defined by
\begin{equation}
\langle f,f'\rangle_{KG}=i\int_\Sigma \Big(f^\ast\partial_\mu f'-f'\partial_\mu f^\ast\Big)\sqrt{h}n^\mu d^{3}X,
\end{equation}
where $h$ is the determinant of the induced metric on a arbitrary space-like surface $\Sigma$ and $n^\mu$ is
the forward directed unit vector normal to $\Sigma$.}. At the large $\tau$ the behavior of the solution controls by the second kind of the Legendre function where can be used to construct the $in$ and $out$ vacuum states for positive and negative frequencies, so that one has
\begin{equation}
\phi(\tau)\approx \frac{\cal N}{\cosh \tau}Q^{-L}_{-\frac{1}{2}-i\nu}(i\sinh\tau).
\end{equation} 
Now the particle $in$ solution, $\phi^{in\,+}(\tau)$, which is the positive frequency solution that behaves as $e^{-i\nu \tau}$ as $\tau\rightarrow-\infty$, and the corresponding anti-particle solution, $\phi^{in\,-}(\tau)$, which behaves as $e^{i\nu\tau}$ as $\tau\rightarrow-\infty$ can indeed build a complete set of solution. Similarly one can define another set built on $out$ state solutions, noting that the particle $out$ state $\phi^{out\,+}(\tau)$ is the positive frequency solution that behaves as $e^{-i\nu\tau}$ at the future infinity. Now each set of basis can be written in terms of others, that means 
\begin{eqnarray}
\phi^{in/out\,+}(\tau)={\cal A}^{in/out}\varphi(\tau)+{\cal B}^{in/out}\varphi^\ast(\tau).
\end{eqnarray}
In our case ${\cal B}\neq0$ and the Bololuibov coefficients ${\cal A}$ and ${\cal B}$ are found as
\begin{eqnarray}
{\cal A}^{in}=e^{-i\frac{\pi}{2}(L-2i\nu)}\frac{1}{\sqrt{2\sinh\pi\nu}},\nonumber\\
{\cal B}^{in}=ie^{i\frac{\pi}{2}(L-2i\nu)}\frac{1}{\sqrt{2\sinh\pi\nu}},
\end{eqnarray}
this means that there is a mixture of states at the past infinity, one can say that, although the modes are globally defined, but the corresponding vacuum could not be a true vacuum for all inertial observers.
\setcounter{equation}{0}
\section{Conclusions}

In this work we had a look at the old problem of instability of de Sitter space to particle creation. As a matter of fact the $O(1,4)$ transformation may rotate the solutions into a linear combination of the positive and negative ones. This means that based only on symmetry group of de Sitter space, a separation into particle and anti-particle is impossible. This can actually be understood by a closely related case of particle creation in a constant uniform electric field, where de Sitter invariant state which, analogous to a time symmetric one, is not a stable vacuum state. We employed the de Sitter plane wave method to study the issue of instability of de Sitter space. In this method the group theoretical
approach is utilized, so that the de Sitter invariance would be manifest, moreover, this approach could
potentially generalize to higher spins and dimensions. We showed that there is a mixture of particle and anti-particles at the past infinity it means that the vacuum constructed from such modes  also unstable to particle creation. The so-called plane wave could potentially define globally in de Sitter space and give rise to the thermal interpretation for the Bunch-Davies or Euclidean vacuum in de Sitter space. There is no antipodal singularity in the plane wave method and also the de Sitter invariance becomes manifest and at the zero-curvature limit, these modes exactly coincide to the usual plane wave in Minkowski space. Moreover, this method can be generalized to study higher spin staff in de Sitter space in a rather simple way, particularly, we discussed on the corresponding modes of massive vector and tensor fields in which the corresponding modes are written in terms of the plane waves. 

Similar procedure of plane wave is used in the context of dS$_4$/CFT$_3$ correspondence. In the embedding space extension the bulk-to-boundary propagator is just proportional to $(\xi\cdot x)^{-h}$ where $h$ is the conformal dimension of the dual operator, $\xi$ is a null vector in the embedding space and $x$ is a point on dS. This formalism has been discussed in several papers (for the application of this extension see \cite{Penedones:2010ue} and Ref.s therein). For every spin-zero primary CFT$_3$ operator such as ${\cal O}$ of conformal weight $h$, there is a bulk scalar field $\Phi(x)$ with mass $m^2=h(3-h)$, where in de Sitter it is not difficult to see that $h=-\sigma$. Generalization of this procedure to de Sitter space would be useful in the sense of better understanding of dS/CFT correspondence.

{\subsection*{Acknowledgements}

We would like to acknowledge M. Alishahiha and M. V. Takook for very useful discussions. We thank  M. Mohsendadeh, E. Yusofi, M. Reza Mozaffar and Ali Mollabashi.  We would also like to thank the referee for particularly helpful comments,
which led to a substantial improvement in the paper. 

\enlargethispage{\baselineskip}

{\subsection*{Appendix: }\label{App. A}

\subsection*{Some useful formulae:} 
In obtaining the Bogoliubov coefficients the following relations have been used
\begin{eqnarray}
Q_a^b(z)&=&\frac{e^{b\pi i}}{2}\{\frac{\Gamma(a+b+1)\Gamma(-b)}{\Gamma(a-b+1)}(\frac{z-1}{z+1})^{b/2}F(-a,a+1,1+b;\frac{1-z}{2})\nonumber\\ 
&&+\Gamma(b)(\frac{z-1}{z+1})^{-b/2}F(-a,a+1,1-b;\frac{1-z}{2})\},\nonumber\\
F(a,b,c;z)&=&\frac{\Gamma(c)\Gamma(c-a-b)}{\Gamma(c-a)\Gamma(c-b)}F(a,b,1+a+b-c;1-z)\nonumber\\
&&+\frac{\Gamma(c)\Gamma(a+b-c)}{\Gamma(a)\Gamma(b)}(1-z)^{c-a-b}F(c-a,c-b,c-a-b+1;1-z),\nonumber\\
F(a,b,c;z)&=&(1-z)^{c-a-b}F(c-a,c-b,c;z).\nonumber
\end{eqnarray}

\subsection*{Flat limit of Plane waves}
A convenient choice for the global coordinates is given by
\begin{equation}
x(X)=H^{-1}\Big(\sinh Ht,\,\,\cosh Ht\sin Hx \,\,\textbf{u}_x,\,\,\cosh Ht\cos Hx\Big),
\end{equation}
 where $\textbf{u}_x=\frac{\textbf{x}}{x}$  is the unit vector along the $\textbf{x}$-direction. Suppose the null vector $\xi=\frac{1}{m}(\sqrt{m^2+\textbf{k}^2},\textbf{k}, \pm m)$, so that the plane wave can be written as 
 \begin{equation}
 (H x\cdot \xi)^{\sigma_\pm}=
\Big(\frac{1}{m}[\sqrt{k^2+m^2}\sinh Ht-\textbf{k}\cdot\textbf{u}_x\cosh Ht\sin Hx\pm m\cosh Ht\cos Hx]\Big)^{\sigma_\pm}
 \end{equation}
and the zero-curvature limit is 
 \begin{equation}
\begin{array}{l}
\lim_{H\rightarrow0}\Big(\pm1+\frac{Ht\sqrt{m^2+k^2}-H\textbf{k}\cdot\textbf{x}}{m}\Big)^{\sigma_\pm}=\\
\lim_{H\rightarrow0}\exp\Big[\pm(-\frac{3}{2}\pm i\frac{m}{H})\frac{H(t\sqrt{m^2+k^2}-\textbf{x}\cdot\textbf{k})}{m}\Big]=\\
\exp[\pm i(t\sqrt{m^2+k^2}-\textbf{x}\cdot\textbf{k})].
\label{limit}
\end{array}
\end{equation}
Note that $\xi= \frac{1}{m}(k^0,\textbf{k},\pm m)$ can be regarded as a wave vector of a Minkowskian particle of mass $m$ and also we have used $\nu=\frac{m}{H}$ (or equivalently, $M_H^2=m^2+(\frac{3}{2}H)^2$). Thus one has 
\begin{equation}
\lim_{H\rightarrow 0} (H x\cdot \xi)^{\sigma_\pm}\longrightarrow e^{\pm ik_\mu X^\mu}.
\end{equation} 
 
 \subsection*{Field equations and their asymptotic behaviors:}

The induced metric on de Sitter hyperboloid is
\begin{equation}
ds^2=d\tau^2-(H^{-1}\cosh H\tau)^2 d\Omega_3^2,\hspace{3mm} d\Omega^2_3=\sum_{i=1}^4(d\omega^i)^2,
\end{equation}
by replacing $ \tan\rho=\sinh H\tau$, where $-\frac{\pi}{2}<\rho<\frac{\pi}{2}$, one obtains:
\begin{equation}\label{newmetric}
ds^2=\frac{1}{H^2\cos^2\rho}(d\rho^2-d\Omega^2_3),
\end{equation}
that is the conformal metric. The Klein-Gordon equation in this metric can be solved by the separation of variables as $\phi(X)=\chi_L(\tau)Y_{Lj}(\Omega)$, one obtains:
\begin{equation}\label{tau}
H^2\ddot{\chi}_L+3H\tanh H\tau \dot\chi_L+\Big(M_H^2+\frac{L(L+2)}{\cosh^2H\tau}\Big)\chi_L=0
\end{equation}
where dot is the derivative with respect to $\tau$, and $L(L+2)$ is the angular part in $S^3$ comes form $(\nabla^2+L(L+2))Y_{Lj}(\omega)=0$. The second differential operator in conformal metric becomes:
\begin{equation}
\chi''(\rho)+2\tan\rho\chi'(\rho)+\Big(L(L+2)+H^{2}M^2_H\Big)\chi(\rho)=0,
\end{equation}  
where prime is the $\rho$ derivative. 

The solutions are given in terms of the hypergeometric functions but the asymptotic behaviors at ${\cal I}^\pm$ can be easily obtained by $\tau\rightarrow\pm\infty$ in (\ref{tau})
\begin{equation}
\ddot{\chi}_L\pm3\dot{\chi}_L+M_H^2\chi_L=0,
\end{equation}
and the asymptotic behaviors are
\begin{equation}
\lim_{t\rightarrow-\infty}\chi_L\sim e^{-\sigma_\pm H\tau},\hspace{6mm} \lim_{t\rightarrow\infty}\chi_L\sim e^{\sigma_\pm H\tau},
\end{equation}
or they can be written as
\begin{equation}
\begin{array}{l}
\chi_L(+\infty)\sim e^{-\frac{3}{2}Ht}\exp\Big[\mp i mt\Big],\\
\chi_L(-\infty)\sim e^{\frac{3}{2}Ht}\exp\Big[\pm i mt\Big],
\end{array}
\end{equation}
noting that $m^2=M_H^2-\frac{9}{4}H^2$. These past and future modes when $H\rightarrow 0$ coincide to what obtains in (\ref{limit}) as the flat limit of the de Sitter plane wave. 

\subsection*{Plane waves in global coordinates}

Here, we want to write the plane wave in terms of the intrinsic hyperboloid de Sitter space in the global coordinates \cite{Epstein:2014jaa, Gazeau:2010mn}. Let us use $\xi=(\xi^0,\pmb{\xi})$ in which $\pmb{\xi} = \xi^0 \textbf{u}_\xi \in \mathbb{R}^4$, $\textbf{u}_\xi \in S^3$, and $\vert \xi^0 \vert = \Vert \pmb{\xi} \Vert$. In the conformal metric (\ref{newmetric}), the $Hx\cdot \xi$ takes the following form 
\begin{gather}
Hx\cdot \xi  = (\tan{\rho})\,\xi^0 - \frac{1}{\cos{\rho}} \textbf{u}_x\cdot\pmb{\xi}
 = \frac{\xi^0 e^{i\rho}}{2i\,\cos{\rho}}\big(1 + z^2 - 2zt\big)\\
\nonumber  \phantom{Hx\cdot \xi  =}{} \mbox{with}\ z = i e^{- i\rho}\mathrm{sgn}\,{\xi^0}, \qquad t \equiv \textbf{u}_x\cdot \textbf{u}_\xi.
\end{gather}
This can be expanded as follows
\begin{gather}
\label{expgegen}
(Hx\cdot\xi)^{\sigma} =  \left(\frac{\xi^0 e^{i\rho}}{2i\,\cos{\rho}}\right)^{\sigma} \big(1 + z^2 - 2zt\big)^{\sigma}= \left(\frac{\xi^0 e^{i\rho}}{2i\,\cos{\rho}}\right)^{\sigma}  \sum_{n=0}^{\infty} z^n  C_n^{-\sigma}(t) , \qquad \Re{\sigma} < \frac{1}{2},
\end{gather}
where  the generating function for Gegenbauer polynomials has been used
\begin{gather}
\label{expgeg}
\big(1 + z^2 - 2 z t\big)^{-\lambda} = \sum_{n=0}^{\infty} z^n \, C_n^{\lambda}(t), \qquad \vert z \vert <1.
\end{gather}
The expansion (\ref{expgegen}) is actually not valid because in our case $\vert z \vert = 1$. The convergence can be obtained by giving a negative imaginary part to the angle $\rho$ which can be achieved by going to the forward tube ${\cal T}^+=\R^5-iV^+\cap X_H^{(c)}$. After making use of the following relations on the Gegenbauer polynomials and $S^3$ normalized hyperspherical harmonics:
\begin{gather}
  C_n^{\lambda}(t)  =
  \frac{1}{\Gamma(\lambda)\Gamma(\lambda-1)} \sum_{k = 0}^{\lfloor \frac{n}{2}\rfloor} c_k   C_{n-2k}^{1} (t) , \nonumber\\
  c_k = \frac{(n-2k + 1) \Gamma(k+\lambda - 1) \Gamma(\lambda + n - k)}{k! \Gamma(n-k + 2)}   ,\label{gegen}\\
 \label{geghyp}  C_L^1(v\cdot v') =    \frac{2\pi^2}{L+1}\sum_{lm} \mathrm{Y}_{Llm}(v) \mathrm{Y}^{\ast}_{Llm}(v') , \qquad v, v' \in S^3,\\
 \mathrm{Y}_{Llm}(u)
=\left(\frac{(L+1)(2l+1)(L-l)!}{2\pi^2(L+l+1)!}\right)^{\frac{1}{2}}
2^ll!\left(\sin\alpha\right)^lC_{L-l}^{l+1}\left(\cos\alpha\right)
Y_{lm}(\theta,\phi) ,
\end{gather}
where $\mathrm{Y}_{Llm}(u)$ is the $S^3$ hyperspherical harmonics for $(L,l,m)\in\N\times\N\times\Z$ with $0\leq l\leq L$ and $-l\leq
m\leq l$, noting that $Y_{lm}$'s are ordinary spherical harmonics:
 \[
Y_{lm}(\theta,\phi)=
(-1)^m\left(\frac{(l-m)!}{(l+m)!}\right)^{\frac{1}{2}}
P_l^m(\cos\theta)e^{im\phi},
\] 
where the $P_l^m$'s are the associated Legendre functions, one obtains
\begin{gather}
\label{geghyp1}
\big(1 + z^2 - 2 z v\cdot v'\big)^{-\lambda} = 2\pi^2 \sum_{Llm} z^{L}P^{\lambda}_{L}\big(z^2\big)\mathrm{Y}_{Llm}(v) \mathrm{Y}^{\ast}_{Llm}(v'),
\end{gather}
where
\begin{gather*}
	P^{\lambda}_{L}\big(z^2\big)   =
	\frac{1}{(L+1)!}\frac{\Gamma(\lambda+L)}{\Gamma(\lambda)}
	{}_2F_1\big(L+\lambda, \lambda - 1 ; L + 2 ; z^2\big)
\end{gather*}
and the integral representation,
\begin{gather*}
z^{L}P^{\lambda}_{L}\big(z^2\big)  \mathrm{Y}_{Llm}(v) = \frac{1}{2\pi^2} \int_{S^3} \big(1 + z^2 - 2 z v\cdot v'\big)^{-\lambda} \mathrm{Y}_{Llm}(v') d\mu(v') .
\end{gather*}

Combining the results of equations~(\ref{expgegen}) and (\ref{geghyp1}) with $\lambda = -\sigma$, leads one to write 

\begin{equation}
(Hx\cdot\xi)^\sigma=2\pi^2\sum_{Llm}\Phi_{Llm}^{\sigma} (x)(\xi^0)^\sigma(\mbox{sgn} \xi^0)^L\mathrm{Y}^\ast_{Llm}(u)
\end{equation}
where \begin{gather}
\nonumber \Phi_{Llm}^{\sigma} (x)= \frac{i^{L-\sigma} e^{-i(L-\sigma)\rho}}{(2\cos{\rho})^{\sigma}} P^{-\sigma}_{L}\big({-}e^{- 2i\rho}\big) \mathrm{Y}_{Llm}(u)\\
\phantom{\Phi_{Llm}^{\sigma} (x)}{}
= \frac{i^{L-\sigma} e^{-i(L-\sigma)\rho}}{(2\cos{\rho})^{\sigma}}\frac{\Gamma(L-\sigma)}{(L+1)!\Gamma(-\sigma)}\, {}_2F_1\big(L-\sigma,-\sigma - 1; L+2;-e^{- 2i\rho}\big) \mathrm{Y}_{Llm}(u) .\end{gather}

\end{document}